\documentclass[preprint,12pt]{elsarticle}

\usepackage{hyperref}
\usepackage{ifpdf}
\usepackage{graphicx}
\usepackage{bm}
\usepackage[english]{babel}
\usepackage[latin1]{inputenc}
\usepackage{multirow}
\usepackage{wrapfig}
\usepackage[svgnames,dvipsnames]{pstricks}
\usepackage{pst-bar}
\usepackage{pst-plot}
\usepackage{pstricks-add}
\usepackage{filecontents}
\usepackage{psfrag}
\usepackage{amsmath}
\usepackage{natbib}

\journal{Comptes Rendues}

\begin{document}

\newcommand{\ket}[1]{|#1\rangle}
\newcommand{\outbracket}[2]{\langle #1 | #2 \rangle}
\newcommand{\inbracket}[2]{| #1 \rangle \langle #2|}
\newcommand{\abs}[1]{\left| #1 \right|}

\begin{frontmatter}



\title{Polarization Control for Slow and Fast Light in Fiber Optical, Raman-assisted, Parametric Amplification}


\author{M. Santagiustina, L. Schenato, C.G. Someda}

\address{Department of Information Engineering, University of Padova, 
via Gradenigo 6/B, 35131 Padova, Italy}

\begin{abstract}
Efficient slow and fast light fiber devices based on narrow band optical parametric
amplification require a strict polarization control of the waves involved
in the interaction. The use of high birefringence and spun fibers is
studied theoretically, possible impairments evaluated, and design parameters determined.
\end{abstract}


\end{frontmatter}


\section{Introduction}
\label{intro}

It has been demonstrated that narrowband optical parametric 
amplification (NBOPA) is a superb technique for inducing slow
and fast light (SFL) in optical fibers \cite{DAH05OE}.  
Large group delay tuning, over selectable, wide frequency bands, 
makes this technique a very promising candidate for many
envisaged applications. Record tunable SFL delays of 
communication digital data signals were achieved 
in dispersion shifted fibers (DSF) \cite{DAH05OE}. In particular,
experimental demonstrations were performed for 10 Gbit/s and 
theoretical predictions for 40 Gbit/s digital signals 
have been given \cite{SHU06OE,SHU06ecoc}.

Theoretical studies resulted into a good understanding of the NBOPA, SFL process
and its intrinsic limitations \cite{DAH05OE, SAN08ieeewt} under the key assumption
of an ideal homogeneous, isotropic fiber. However, real fibers are not likely
to be homogeneous nor isotropic, and NBOPA gain and delay are affected
by the longitudinal variations of the fiber linear propagation parameters 
such as the zero dispersion wavelength (ZDW) fluctuations \cite{SHU07ofc} 
and the random birefrigence \cite{SAN07ieeeps}.

Fluctuations of the ZDW cause gain broadening and thus a decrease in the achieved delay. 
The main contribution to ZDW shifts comes from changes in the fiber core effective area 
occurring during the drawing process \cite{EIS97JLT,KAR98JOSAB}, so 
ZDW shift is not time varying. Eventually, the ZDW shift can be measured \cite{SHU07ofc,EIS97JLT}
and the gain broadening can be highly mitigated by selecting uniform samples, and by increasing the pump power.

The polarization sensitivity of optical parametric interactions 
is a much more intriguing issue \cite{MAR03JOSAB,LIN04JOSAB}:
the parametric gain coefficient is maximum when the pump and the signal have the same state of polarization
(SOP), while it vanishes for orthogonal SOPs.
So, maintaining a strict control of the pump and signal relative SOPs along the fiber
is a critical issue for attaining reliable SFL devices based on NBOPA.

The most widespread fibers (i.e. telecommunication ones) are not isotropic,
though their birefringence is very low (here, they will be referred to as low birefringence - LoBi - fibers).
The residual birefringence stems from the manufacturing imperfections (e.g. small asymmetries of the 
fiber core circular section) and from the fiber operating conditions 
(e.g. stresses, bending, twisting and temperature changes) due to environmental conditions;
all these factors eventually break the polarization degeneracy of the fundamental
mode, and the fiber becomes birefringent.
The residual, stochastic, low birefringence causes a random phenomenology known as 
polarization mode dispersion (PMD) \cite{FOS91JLT, GOR00PNAS}.
The parametric interaction is very sensitive to PMD, that modifies
the pump, signal and idler SOPs in a random fashion along the fiber \cite{LIN04OL}.
Detailed analyses of the effects of PMD on NBOPA gain and SFL delay can be found in
refs. \cite{WIL08JLT, SHU08book, SCH08JLT}. The effect is very pronounced and harmful because: 
a) it increases with the signal-pump frequency detuning, which is very large for NBOPA 
\cite{SCH08ofc}; b) differently from ZDW fluctuations, which are deterministic, 
PMD is a random, time varying phenomenon \cite{FOS91JLT,GOR00PNAS}.

In this contribution, two special fiber types that can be used to control the polarization
of waves interacting in a NBOPA, SFL device will be theoretically and numerically
studied. The paper aims at presenting fundamental design information for improving the
NBOPA, SFL fiber devices beyond the present state of the art.
High birefringence (HiBi) fibers, in which the SOP is maintained
\cite{PAR87AO}, are an obvious choice to mimic the isotropic ideal case, but not the 
only one. Significant reductions of random polarization effects in Raman \cite{BET08PTL}, 
Brillouin \cite{GAL08PTL} and parametric \cite{FER09ofc,SAN09PTL} amplifiers, have been recently predicted 
when unidirectionally spun (US) fibers are considered. 
Here, the weight of possible negative effects in HiBi fibers and the positive effects of unidirectional spinning
for on NBOPA-SFL are quantified. 

The paper is organized as follows. In section \ref{vecth} the equations describing the 
NBOPA with polarized fields, in different fiber types, are introduced. The main features 
of propagation in LoBi, HiBi and US fibers will be also recalled in this section.
In section \ref{LoBi} the performance in LoBi fibers will be assessed.
The study of NBOPA, SFL in HiBi and US fibers will be carried out in sections 
\ref{HiBi} and \ref{us}, respectively. Finally, conclusions will be drawn in section \ref{conc}.

\section{Propagation model}
\label{vecth}

The aim of this section is to provide a unified model to describe the NBOPA 
propagation for different fiber types, e.g. LoBi, HiBi and US. Moreover, the main features 
of the propagation of optical signals in such fiber types will be also reviewed.

Let us define $\ket{A_p(z)}, \ket{A_s(z,t)}, \ket{A_i(z,t)}$, the Jones vectors of 
the pump, signal and idler waves respectively. For detailed definitions of the ket 
$\ket{}$ symbol and the bracket operators $\inbracket{}{}, \outbracket{}{}$,
one can refer to \cite{GOR00PNAS}.

In the undepleted pump approximation, neglecting the nonlinear effects of 
the signal and idler on the pump and by considering a continuous wave pump,
the equations governing the nonlinear interaction of the slowly varying envelopes
for the pump, signal and idler read
\cite{WIL08JLT, SHU08book, SCH08JLT, LIN04JLT, TRIL94JOSAB}:
\begin{equation}
\begin{split}
&\frac{d \ket{A_p}}{dz} = \left[ {\cal L}_p + {\cal S}_p \left( \ket{A_p} \right) \right] \ket{A_p}, \\
& \frac{\partial \ket{A_s}}{\partial z} = \left[ {\cal L}_s + {\cal X}_s \left( \ket{A_p} \right) 
+ {\cal R}_s \left( \ket{A_p} \right) \right] \ket{A_s} + {\cal F}_s \left( \ket{A_p} \right) \; \ket{A_i^{\ast}}\\
& \frac{\partial \ket{A_i}}{\partial z} = \left[ {\cal L}_i + {\cal X}_i \left( \ket{A_p} \right) + 
{\cal R}_i \left( \ket{A_p} \right) \right] \ket{A_i} + {\cal F}_i \left( \ket{A_p} \right) \; \ket{A_s^{\ast}}.
\end{split}\label{fopaeq}
\end{equation}
The operators ${\cal L}_{p,s,i}$ account for the linear propagation properties, while ${\cal S}_p,
{\cal X}_{s,i}, {\cal R}_{s,i}, {\cal F}_{s,i}$ are operators that depend on the pump wave
$\ket{A_p}$, and take into account the nonlinear effects that are relevant for each wave, i.e.
self-phase modulation, cross-phase modulation, Raman scattering and four-wave mixing. 
The linear operators are described in this section, as the different fiber type
(LoBi, HiBi, US) properties are dictated by these operators. Nonlinear operators present very complicated
structures, that are described in the Appendix \ref{appendice}, for the sake of completeness.

The linear operators are defined as:
\begin{equation}
\label{linear}
{\cal L}_h= - \alpha_h + j \beta_h - \beta_{1h} \frac{\partial}{\partial t} 
- j \frac{1}{2} \bar{\beta}(\omega_h) \cdot \bar{\sigma} + \frac{1}{2} \bar{\delta_h} \cdot \bar{\sigma}
\frac{\partial}{\partial t}, \; h=p,s,i.
\end{equation}

In eqs. \ref{linear}, $\alpha_{p,s,i}$ are the loss coefficients, $\beta_{p,s,i}=\beta(\omega_{p,s,i})$
the mean wavenumbers at the optical angular frequencies of the pump, signal and idler $\omega_{p,s,i}=2 \pi c_0/\lambda_{p,s,i}$ 
satisfying $2\omega_p=\omega_s+\omega_i$. 
The nonlinear phase matching condition to be satisfied is:
$2\beta_p - \beta_s -\beta_i =\Delta \beta= \beta_{2p} \, (\omega_s-\omega_p)^2+ \beta_{4p} \, (\omega_s-\omega_p)^4/12= -2 \gamma P_0$,
where $\beta_{np}$ is the $n$-th derivative of
$\beta(\omega)$ with respect to the angular frequency, calculated at $\omega_p$ \cite{MAR96OL}, and
$P_0=\outbracket{A_p(0)}{A_p(0)}$ is the input pump power. 
The parameters used in the following simulations are: ZDW, $\lambda_0=1.5423~\mu$m; $\lambda_p=1.53~\mu$m; 
$\beta_{2p} \simeq \beta_{30}(\omega_p-\omega_0)$; $\beta_{30}=1.14 \cdot 10^{-40}~\mathrm{s}^3/\mathrm{m}$; 
$\beta_{4p}=-5 \cdot 10^{-55}~\mathrm{s}^4/\mathrm{m}$; 
$P_0=1 \div 5$~W. Then, the NBOPA phase matched wavelengths are: $\lambda_s \simeq 1.3927~\mu$m, 
$\lambda_i \simeq 1.7096~\mu$m.

If the reference frame ($z,t$) used in eqs. \ref{fopaeq} is travelling at the signal group velocity 
$v_g(\omega_s)$, one also gets $\beta_{1h}=0$, for the signal ($h=s$) and 
$\beta_{1h}=1/v_g(\omega_s)-1/v_g(\omega_i)$ for the idler ($h=i$).

The effects of fiber birefringence are accounted for by the last two terms of the operator (\ref{linear}).
In the first term the Stokes vector $\bar{\beta}(z,\omega)$ is the local birefringence vector that
describes the birefringence at each point within the fiber; $\bar{\sigma}$ is the
vector of the Pauli spin matrices \cite{GOR00PNAS}. 
In the second term: $\bar{\delta_h}= \partial \bar{\beta}/ \partial \omega$ calculated
at $\omega_h$ \cite{LIN04JLT}.

Results for an ideal, isotropic fiber, can be obtained by setting $\bar{\beta}=0 \; \forall z, \; \forall \omega$.
In real fibers, the properties of the birefringence vector, which change
from one type to the other, are very important. For this reason, they need
to be specified in detail.

For LoBi (unspun) fibers $\bar{\beta}(z,\omega)=\bar{\beta}_{un}(z,\omega)= [\beta_1, \beta_2, 0]$ is a random
vector. Its evolution along $z$ can be obtained by means of the
so-called random modulus model (RMM) \cite{WAI96JLT}, i.e. its components are generated by the
following Langevin equations:
\begin{equation}
\frac{d\beta_{i}}{dz}=-\rho \beta_{i}+\nu \eta_{i},\ \ \ \ \ \ \ \ i=1,2 \;
\label{rmm}
\end{equation}
where $\beta_{i} \; (i=1,2)$ are Gaussian stochastic variables of zero mean and variance 
$\nu^{2}_{\beta}=\nu^{2}/(2\rho)$ and $\eta_{i}(z)$ for $i=1,2$ are independent, 
Gaussian white noises of zero mean and unitary variance.
Note that $\beta_3(z) \equiv 0$ is set in the numerical solutions, as is commonly 
assumed \cite{WAI96JLT} and experimentally verified in most cases \cite{GAL01OL}.
This condition means that the fiber does not exhibit any circular
birefringence.

In the RMM, PMD is actually described by two length scales: the beat length 
$L_B=2\pi / (\sqrt{2} \nu_{\beta})$, and the birefringence correlation length 
$L_F=1/\rho $. The former depends on the frequency ($L_B(\omega)=\omega_0 L_B(\omega_0)/\omega$)
and describes the length scale of polarization changes, while the latter is frequency independent
($L_F=9$~m in our simulations), and accounts for the length scale of birefringence
changes.
Both lengths contribute to determine the PMD coefficient \cite{WAI96JLT},
hereinafter defined as $D=\sqrt{\langle \Delta \tau^2 \rangle/L}$, where
$\langle \Delta \tau^2 \rangle$ is the fiber mean square differential group delay 
(DGD), and $L$ is the fiber length. 
The DGD is defined as the time delay between pulses, at the same carrier frequency,
launched along the two principal states of polarization (PSPs) \cite{WAG86} in LoBi fibers, and along
the birefringence axes in HiBi fibers. The PSPs are defined as those input SOPs whose corresponding output SOPs
are frequency independent at first order \cite{WAG86}.
The typical coefficient $D$, in LoBi fibers, ranges from $10^{-2}~\mathrm{ps}/\sqrt{\mathrm{km}}$, for low PMD fibers,
to more than $10^{-1}~\mathrm{ps}/\sqrt{\mathrm{km}}$ for high PMD ones. 

The second type of fibers, HiBi, are for example realized by inducing internal stresses 
(PANDA fibers) or by making an asymmetric core (elliptical core fibers). 
In both cases the large intrinsic guide birefringence, voluntarily introduced during the manufacturing
process, dominates over random effects. Then, in HiBi fibers the linear birefringence vector 
$\bar{\beta}$ has a predominant, deterministic, linear contribution $\bar{\beta}_{HiBi}$ whose modulus is related  
to the DGD $\Delta \tau$ by $\abs{\bar{\beta}_{HiBi}}=c_0 \Delta \tau / (\lambda L)$. 
Input SOPs parallel to birefringence axes ($\pm \hat{\beta}_{HiBi}=\pm \bar{\beta}_{HiBi} / \abs{\bar{\beta}_{HiBi}}$)
are polarization eigenstates, and so they travel unchanged through the entire fiber length.
For this reason HiBi fibers are also referred to as polarization maintaining (PM) ones.
Though the deterministic birefringence is overwhelming, random mode coupling along the fiber still
exists. Furthermore, input misalignment may also result in a nonvanishing cross-polarized orthogonal SOP.
The tolerance of the SFL technique with respect to SOP misalignement has been investigated
and will be discussedin the following.
Moreover, for a more realistic simulation of the unwanted effects, a random component, described
again by the RMM, has been added to the deterministic part of the birefringence
$\bar{\beta}=\bar{\beta}_{HiBi}+\bar{\beta}_{ran}$ \cite{FOS91JLT}. The value of $D$ has
been chosen to yield the typical polarization cross-talk ratio (PXR) \cite{NOD86JLT}
of commercial HiBi fiber, always better than $20$ dB.
As for the effects of input misalignments, the delay was calculated
numerically exploring all possible linear input SOPs.

Finally, US fibers have been considered; fiber spinning is a manufacturing process routinely 
performed while drawing fibers from LoBi preforms.
In particular, periodic spinning functions, in which the fiber is turned 
alternatively clokwise and counterclockwise, are often
applied to reduce external stresses and fiber PMD \cite{BAR81AO}.
Unidirectional spinning, besides reducing the DGD, has 
been predicted to enhance the SOP alignment of optical signals at different frequencies, in particular in nonlinear 
fiber amplifiers \cite{BET08PTL,GAL08PTL,FER09ofc,SAN09PTL}.
The birefringence vector of a spun fiber can be obtained from that of the unspun case through 
the transformation
$\bar{\beta}(z,\omega) = R_3[2\phi(z)] \bar{\beta}_{un}(z,\omega)$ where $R_{3}$ is a Mueller matrix 
representing a rotation around the third axis in the Stokes space ($\hat u_3$). 
For an US fiber the angle of rotation is given by the constant spin function $\phi(z)=2\pi z/p$, 
where $p$ is called the spin pitch. 
It has been shown \cite{PAL06JLT} that when $p^2 \ll L_B^2$, the polarization 
properties of the US fiber can be effectively described by a simplified model (SM). In 
the SM, the fiber can be represented by an equivalent
birefringence vector, with a random linear and a deterministic circular component: 
\begin{equation}
\label{sm}
\bar\beta_{eq}(z) = (\sqrt{2\mu}\,\xi_1(z),\, \sqrt{2\mu}\,\xi_2(z),\,-\chi)^T, 
\end{equation}
where $\xi_i(z)$ are statistically independent Gaussian white noises and the moduli of
the linear and circular components are given by:
\begin{equation}
\label{muegamma}
\mu = \frac{2 L_F ( \pi p )^2}{L_B^2 [p^2 + ( 4\pi L_F)^2]}, \; \; \; \chi = \frac{4 \pi L_F \mu}{p}.
\end{equation}
The results of the SM are recalled here because in section \ref{us} they will
contribute to explain the mitigation of polarization effects .

For LoBi and US fibers, several hundreds statistical realizations of the stochastic processes, 
and subsequent integrations of eqs. \ref{fopaeq}, have been realized to calculate the mean 
gain and time delay.

To conclude this section, let us remark that when propagation in birefringent media
is considered, the group velocity cannot be uniquely defined, as observed by Haus 
\cite{HAU00JOSAB}. Exceptions are represented by the special cases in which the input SOPs 
coincide with the PSPs, in LoBi fibers, or birefringence axes, in HiBi fibers. 
In those two cases, a different value is found for the group velocity for each PSP or axis.
In all other conditions, the standard formula $\Delta T_g=d (\beta z)/d \omega=z/v_g(\omega)$
for the group delay looses its physical significance.
Hence, in the numerical integrations, the group delay $\Delta T_g$ has been
evaluated as the first moment of the pulse as a function of
time \cite{SHU08book,SMI70AJP} (the input pulse was Gaussian, $70$~ps FWHM) 
\begin{equation}
\Delta T_g (z) =\frac{\displaystyle\int{t \; \outbracket{A_s(z,t)}{A_s(z,t)} \; dt}}
{\displaystyle\int{ \outbracket{A_s(z,t)}{A_s(z,t)} \; dt}}
\label{delay}
\end{equation}
In particular, the difference between the arrival time when the SFL pump
is on and when it is off,
($\Delta T = \Delta T_g^{on}-\Delta T_g^{off}$), is calculated according to this definition.

%It is also useful to evaluate the pulse distortion to evaluate the performance of the NBOPA technique in birefringent
%media. This can be performed by calculating: a) the second moment of 
%time deviation around the delayed arrival time:
%\begin{equation}
%\Delta T^{(2)}(z) =\frac{\displaystyle\int{(t-\Delta T)^2 \; \outbracket{A_s}{A_s}} \; dt}{\displaystyle\int{\outbracket{A_s}{A_s}} \; dt};
%\label{dt2}
%\end{equation}
%and b) the distortion parameter
%\begin{equation}
%\delta=\abs{1-\frac{\Delta T^{(2)}(z=L)}{\Delta T^{(2)}(z=0)}}.
%\label{distortion}
%\end{equation}
%which measures the departure of the output pulse shape from the input pulse shape.

\section{Low birefringence fibers}
\label{LoBi}

Let us first briefly review the effects of PMD on SFL propagation in NBOPA \cite{WIL08JLT, SHU08book, SCH08JLT}. 
For LoBi fibers the loss of alignment between signal and pump SOPs causes a reduction of the mean gain,
 which in turn
translates into a delay reduction. For small PMD coefficients 
%($D < 0.05~\mathrm{ps}/\sqrt{\mathrm{km}}$)
the mean delay $\langle \Delta T \rangle$ can be calculated with the ideal isotropic case formula
\cite{SAN08ieeewt,SCH08JLT}, where the gain is replaced by the mean gain, i.e.:
 \begin{equation}
  \begin{split}
\langle \Delta T \rangle=&L \sqrt{6 k_2}\,\sqrt{\frac{1}{L_D}}\,\sqrt{1+\frac{2}{3}
\frac{L_D}{\langle L_{NL} \rangle}}\, \sqrt{1 +\sqrt{1
+\frac{2}{3}\frac{L_D}{\langle L_{NL} \rangle}}}\,\,\,\times\\
&\qquad\times\left[1-\frac{\langle L_{NL} \rangle }{L}\tanh\left(
\frac{L}{\langle L_{NL} \rangle }\right)\right]
  \end{split}\label{eq:DT}   
 \end{equation}
where $L_D=-\beta_{4p}/\beta^2_{2p}$ and the mean nonlinear length $\langle L_{NL} \rangle $ is related to the mean
gain $\langle G \rangle$ by:
\begin{equation}
\label{gain}
\frac{L}{\langle L_{NL} \rangle }=\cosh^{-1}\left(\sqrt{\langle G \rangle}\right).
\end{equation}

For larger random birefringence, strong pulse distortion sets in, and the delay decreases faster 
than the gain. 
Though mean quantities follow the ideal relation, random birefringence causes a large uncertainty in 
the actual delay. This fact can be easily grasped from fig. \ref{figus1}, where
the NBOPA gain versus delay, for many realization of an unspun fiber, for a 
PMD coefficient $D=0.05~\mathrm{ps}/\sqrt{\mathrm{km}}$, is shown and 
compared with the case (squares) for $D=0$ (ideal isotropic fiber).
In the real case, PMD shifts the phase-matching condition \cite{WIL08JLT,SCH08JLT},
and therefore the gain is reduced, while eq. \ref{eq:DT} is still valid.
As we said, the delay standard deviation, represented by vertical red bars, is very large; around $20-30$~ps
i.e. almost 50 percent of the mean achieved delay. Finally, pulses are affected by a severe distortion.
The large fluctuations of the delay are probably the most detrimental impairment
introduced by PMD. So, regardless of LoBi fibers being the most commonly
used ones, using them for making stable NBOPA based SFL devices is highly hampered.

\section{High birefringence fibers}
\label{HiBi}

\begin{figure}
 % This file is generated by the MATLAB m-file laprint.m. It can be included
% into LaTeX documents using the packages graphicx, color and psfrag.
% It is accompanied by a postscript file. A sample LaTeX file is:
%    \documentclass{article}\usepackage{graphicx,color,psfrag}
%    \begin{document}\input{figurePM}\end{document}
% See http://www.mathworks.de/matlabcentral/fileexchange/loadFile.do?objectId=4638
% for recent versions of laprint.m.
%
% created by:           LaPrint version 3.16 (13.9.2004)
% created on:           01-Sep-2009 16:32:35
% eps bounding box:     12 cm x 10.3032 cm
% comment:              
%
\begin{psfrags}%
\psfragscanon%
%
% text strings:
\psfrag{s01}[t][t]{\color[rgb]{0,0,0}\setlength{\tabcolsep}{0pt}\begin{tabular}{c}Misalignment~[degrees]\end{tabular}}%
\psfrag{s02}[b][b]{\color[rgb]{0,0,0}\setlength{\tabcolsep}{0pt}\begin{tabular}{c}$\Delta T$~[ps]\end{tabular}}%
\psfrag{s05}[][]{\color[rgb]{0,0,0}\setlength{\tabcolsep}{0pt}\begin{tabular}{c} \end{tabular}}%
\psfrag{s06}[][]{\color[rgb]{0,0,0}\setlength{\tabcolsep}{0pt}\begin{tabular}{c} \end{tabular}}%
%
% xticklabels:
\psfrag{x12}[t][t]{0}%
\psfrag{x13}[t][t]{10}%
\psfrag{x14}[t][t]{20}%
\psfrag{x15}[t][t]{30}%
\psfrag{x16}[t][t]{40}%
\psfrag{x17}[t][t]{50}%
\psfrag{x18}[t][t]{60}%
\psfrag{x19}[t][t]{70}%
\psfrag{x20}[t][t]{80}%
\psfrag{x21}[t][t]{90}%
%
% yticklabels:
\psfrag{v01}[r][r]{0}%
\psfrag{v02}[r][r]{0.1}%
\psfrag{v03}[r][r]{0.2}%
\psfrag{v04}[r][r]{0.3}%
\psfrag{v05}[r][r]{0.4}%
\psfrag{v06}[r][r]{0.5}%
\psfrag{v07}[r][r]{}%
\psfrag{v08}[r][r]{0}%
\psfrag{v09}[r][r]{}%
\psfrag{v10}[r][r]{10}%
\psfrag{v11}[r][r]{}%
\psfrag{v12}[r][r]{20}%
\psfrag{v13}[r][r]{}%
\psfrag{v14}[r][r]{30}%
%
% Figure:
\includegraphics[width=\textwidth]{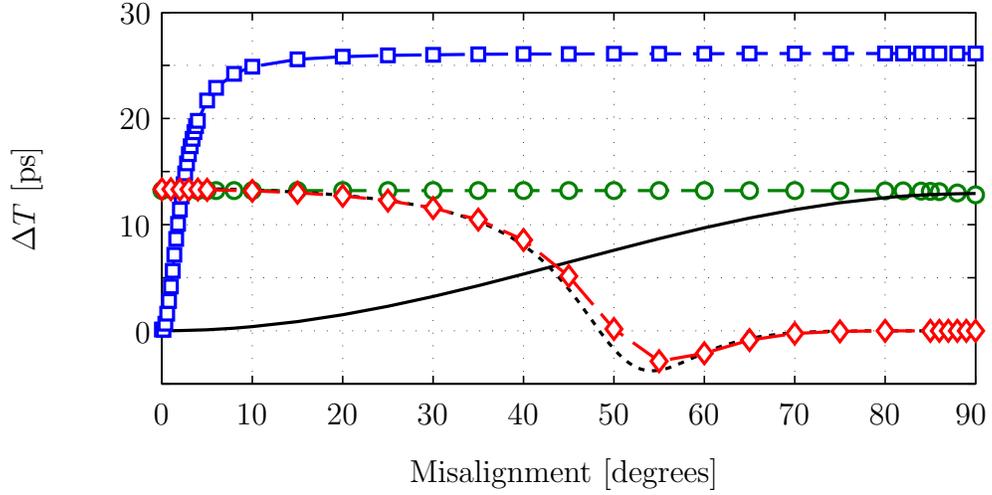}%
\end{psfrags}%
%
% End figurePM.tex

  \caption{Delay vs. input misalignment. Signal - fast axis misalignment: Solid curve, no pump;
Circles, pump aligned with the fast axis; Squares, pump aligned with the slow axis. 
Pump - fast axis misalignment: Diamonds, signal aligned with the fast axis; Dashed curve, analytical result.}
\label{figpmdelay}
\end{figure}

For SFL in HiBi fibers, the key parameter is power splitting among polarizations. In the ideal case, 
since birefringence axes are eigenpolarizations, no cros-polarization
coupling occurs if signal and
pump are co-polarized, and aligned along $\pm \hat{\beta}_{HiBi}$ at the fiber input. Under
real conditions, however, some
coupling always exists because of input misalignment, either of the signal or of the pump, with respect
to the birefringence axes, and/or because of small imperfections in the fiber.
Therefore, the robustness of the SFL scheme against input misalignements has
been tested. The main results are summarized in fig. \ref{figpmdelay}.

The delay was first calculated as a function of the polarization misalignment between the (linear) signal SOP 
and the fast fiber axis ($+ \hat{\beta}_{HiBi}$), along which the pump is suposed to
be launched (empty circles). As we see, the delay is essentially unaffected
for almost all signal SOP misalignements; it decreases significantly only when the signal
is launched almost orthogonally. The explanation of the small change in delay is that a large polarization 
dependent gain is generated in this case \cite{SHU08book}; therefore, the signal polarization
is attracted, by a nonlinear polarization pulling effect similar to that described for Brillouin and Raman amplification
\cite{ZAD08OE,MAR08OE}), towards the direction yielding maximum gain. In practice, the portion of the signal
pulse polarized along the minimum gain direction (slow axis) is weakly amplified, so that the
pulse center of mass actually coincides with that of the powerful pulse
component on the fast axis.
When the pump and signal SOPs are orthogonal (i.e., $90$ degrees misalignment) there is no gain,
and consequently no SFL effect. Hence,
the delay tends to coincide with that due to linear birefringence. This has been verified by propagating
the signal pulse without the pump (solid line); as the misalignment between the signal and the fast axis increases,
more signal power is launched on the slow axis, and delayed because of the different group velocities.
A mirror-like behavior is observed if we launch the pump on the slow axis (squares in fig. \ref{figpmdelay}). 
In this case the delay is the sum of the SFL induced delay and that due to the change in the signal propagation axis.

The effects of the pump input SOP misalignment are shown in fig. \ref{figpmdelay} (empty diamonds).
As more pump power is launched on the slow axis, orthogonally to the signal, the gain,
and consequently the delay, decrease; both tend to zero when the misalignement tends to $90$ degrees.
Remarkably, there are certain input pump SOP's for which the delay becomes negative. This stems
from the fact that the input signal
wavelength is left constant, at the phase matching condition obtained when the pump is launched on the fast
axis. But as the effective pump power decreases because of the misalignement, the phase matching condition
shifts, and consequently the signal spectrum is now in a frequency band
where the delay is negative. This explanation has been verified by
calculating
the delay with the ideal exact formula (i.e. eq. 9 of \cite{SCH08JLT});
the results are presented in fig. \ref{figpmdelay} by the dashed curve.

We may conclude that the delay reduction due to signal and pump input misalignment with respect to fiber axis 
is negligible, if the angle is less than $10$ degrees, an easy condition to
satisfy in practice.

Finally, we considered the random coupling, bearing in mind that this is a
very small effect in good-quality HiBi fibers.
Typically the PXR, i.e. the ratio between power on orthogonal axes at the output of $100$m
of fiber, is better than $20$ dB.
To evaluate these effects of the residual random coupling on the SFL delay,
a stochastic component, obtained
through RMM, was added to the deterministic linear birefringence vector,
so that the total vector becomes:
$\bar{\beta}=\bar{\beta}_{HiBi}+\bar{\beta}_{ran}$.
The value of $D$ was chosen such that the probability of getting realizations with a PXR
in excess of $20$ dB was very low.
The results of a set of statistical realizations of this random process 
%are presented in fig. \ref{figpmdist}. Note 
show that the maximum spread in the time delay is less than 1 percent of the mean value, very close
to the ideal value. We conclude that the effect of the residual random coupling on
SFL delay is negligible in HiBi fibers.

\section{Unidirectionally spun fibers}
\label{us}

\begin{figure}
\begin{center}
  % This file is generated by the MATLAB m-file laprint.m. It can be included
% into LaTeX documents using the packages graphicx, color and psfrag.
% It is accompanied by a postscript file. A sample LaTeX file is:
%    \documentclass{article}\usepackage{graphicx,color,psfrag}
%    \begin{document}\input{figure1}\end{document}
% See http://www.mathworks.de/matlabcentral/fileexchange/loadFile.do?objectId=4638
% for recent versions of laprint.m.
%
% created by:           LaPrint version 3.16 (13.9.2004)
% created on:           27-Aug-2009 16:54:54
% eps bounding box:     15 cm x 7.8516 cm
% comment:              
%
\begin{psfrags}%
\psfragscanon%
%
% text strings:
\psfrag{s01}[t][t]{\color[rgb]{0,0,0}\setlength{\tabcolsep}{0pt}\begin{tabular}{c}Gain~[dB]\end{tabular}}%
\psfrag{s02}[b][b]{\color[rgb]{0,0,0}\setlength{\tabcolsep}{0pt}\begin{tabular}{c}$\Delta T$~[ps]\end{tabular}}%
%
% xticklabels:
\psfrag{x01}[t][t]{0}%
\psfrag{x02}[t][t]{0.1}%
\psfrag{x03}[t][t]{0.2}%
\psfrag{x04}[t][t]{0.3}%
\psfrag{x05}[t][t]{0.4}%
\psfrag{x06}[t][t]{0.5}%
\psfrag{x07}[t][t]{0.6}%
\psfrag{x08}[t][t]{0.7}%
\psfrag{x09}[t][t]{0.8}%
\psfrag{x10}[t][t]{0.9}%
\psfrag{x11}[t][t]{1}%
\psfrag{x12}[t][t]{0}%
\psfrag{x13}[t][t]{20}%
\psfrag{x14}[t][t]{40}%
\psfrag{x15}[t][t]{60}%
%
% yticklabels:
\psfrag{v01}[r][r]{0}%
\psfrag{v02}[r][r]{0.2}%
\psfrag{v03}[r][r]{0.4}%
\psfrag{v04}[r][r]{0.6}%
\psfrag{v05}[r][r]{0.8}%
\psfrag{v06}[r][r]{1}%
\psfrag{v07}[r][r]{0}%
\psfrag{v08}[r][r]{50}%
\psfrag{v09}[r][r]{100}%
\psfrag{a}[][]{\textbf{a}}%
\psfrag{b}[][]{\textbf{b}}%
\psfrag{c}[][]{\textbf{c}}
\psfrag{d}[][]{\textbf{d}}%
% Figure:
\includegraphics[width=\textwidth]{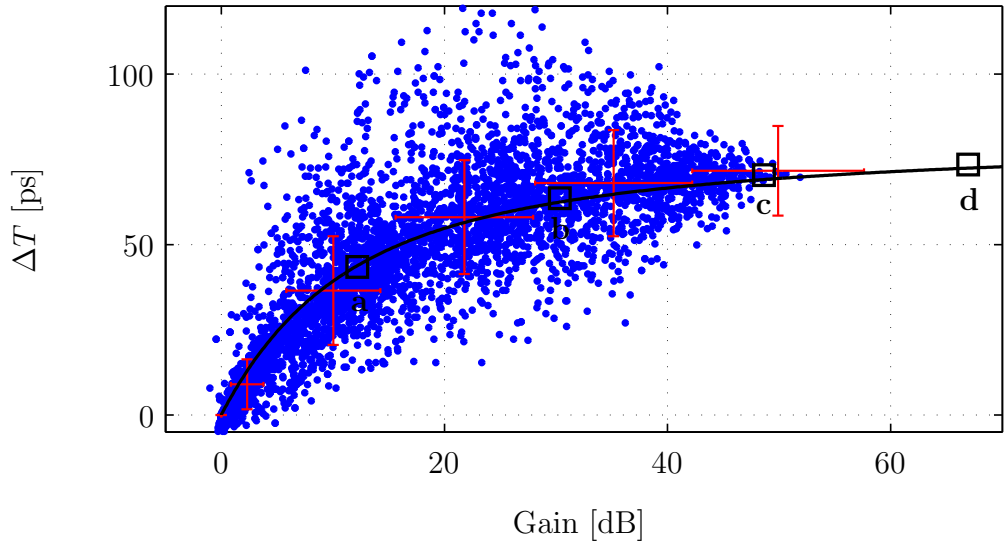}%
\end{psfrags}%
%
% End figure1.tex

  \end{center}
%  \centerline{\includegraphics[width=\textwidth]{figures/delayvsgainun.jpg}}
  \caption{Delay vs. gain from the numerical solutions of eqs. \ref{fopaeq} (L=1km)
for $P_0$: a) 1~W; b) 2~W; c) 3~W; d) 4~W. Squares, isotropic fiber ($D=0$);
dots, LoBi fiber ($D=0.05~\mathrm{ps}/\sqrt{\mathrm{km}}$); solid line, eq.~\ref{eq:DT}. Bars define the gain and
delay standard deviation; their crossing point is the mean value.}
\label{figus1}
\end{figure}

\begin{figure} 
\begin{center}
  % This file is generated by the MATLAB m-file laprint.m. It can be included
% into LaTeX documents using the packages graphicx, color and psfrag.
% It is accompanied by a postscript file. A sample LaTeX file is:
%    \documentclass{article}\usepackage{graphicx,color,psfrag}
%    \begin{document}\input{figure2}\end{document}
% See http://www.mathworks.de/matlabcentral/fileexchange/loadFile.do?objectId=4638
% for recent versions of laprint.m.
%
% created by:           LaPrint version 3.16 (13.9.2004)
% created on:           27-Aug-2009 16:55:01
% eps bounding box:     15 cm x 7.8516 cm
% comment:              
%
\begin{psfrags}%
\psfragscanon%
%
% text strings:
\psfrag{s01}[t][t]{\color[rgb]{0,0,0}\setlength{\tabcolsep}{0pt}\begin{tabular}{c}Gain~[dB]\end{tabular}}%
\psfrag{s02}[b][b]{\color[rgb]{0,0,0}\setlength{\tabcolsep}{0pt}\begin{tabular}{c}$\Delta T$~[ps]\end{tabular}}%
%
% xticklabels:
\psfrag{x01}[t][t]{0}%
\psfrag{x02}[t][t]{0.1}%
\psfrag{x03}[t][t]{0.2}%
\psfrag{x04}[t][t]{0.3}%
\psfrag{x05}[t][t]{0.4}%
\psfrag{x06}[t][t]{0.5}%
\psfrag{x07}[t][t]{0.6}%
\psfrag{x08}[t][t]{0.7}%
\psfrag{x09}[t][t]{0.8}%
\psfrag{x10}[t][t]{0.9}%
\psfrag{x11}[t][t]{1}%
\psfrag{x12}[t][t]{0}%
\psfrag{x13}[t][t]{20}%
\psfrag{x14}[t][t]{40}%
\psfrag{x15}[t][t]{60}%
%
% yticklabels:
\psfrag{v01}[r][r]{0}%
\psfrag{v02}[r][r]{0.2}%
\psfrag{v03}[r][r]{0.4}%
\psfrag{v04}[r][r]{0.6}%
\psfrag{v05}[r][r]{0.8}%
\psfrag{v06}[r][r]{1}%
\psfrag{v07}[r][r]{0}%
\psfrag{v08}[r][r]{50}%
\psfrag{v09}[r][r]{100}%
\psfrag{a}[][]{\textbf{a}}%
\psfrag{b}[][]{\textbf{b}}%
\psfrag{c}[][]{\textbf{c}}
\psfrag{d}[][]{\textbf{d}}%
% Figure:
\includegraphics[width=\textwidth]{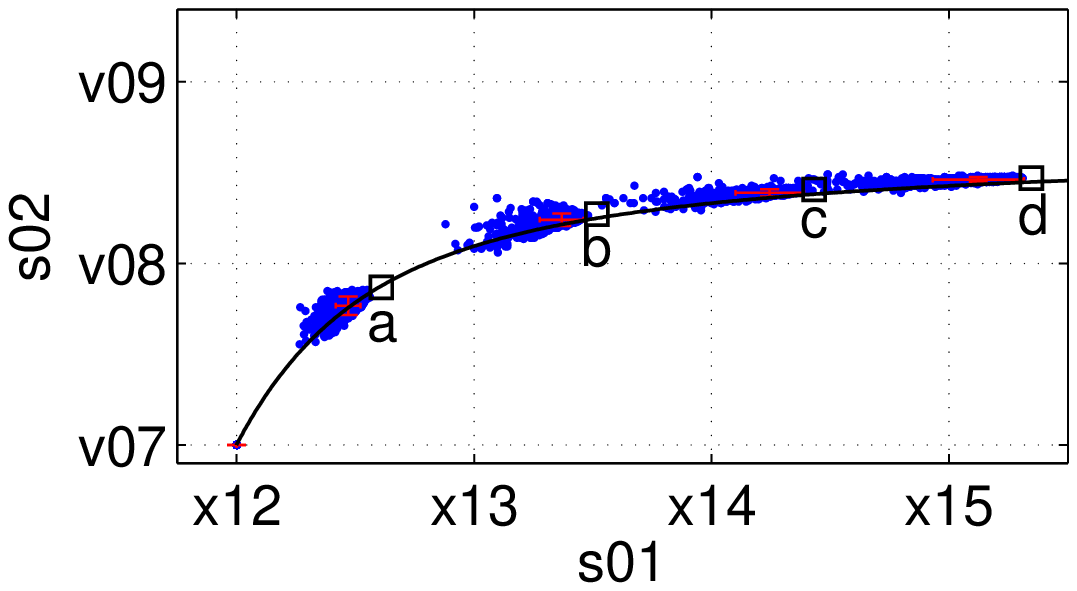}%
\end{psfrags}%
%
% End figure2.tex

  \end{center}
  \caption{Same as fig. \ref{figus1}, but for a US fiber, $p=2 \; m$.}
\label{figus2}
\end{figure}

\begin{figure}
\begin{center}
  % This file is generated by the MATLAB m-file laprint.m. It can be included
% into LaTeX documents using the packages graphicx, color and psfrag.
% It is accompanied by a postscript file. A sample LaTeX file is:
%    \documentclass{article}\usepackage{graphicx,color,psfrag}
%    \begin{document}\input{figure3}\end{document}
% See http://www.mathworks.de/matlabcentral/fileexchange/loadFile.do?objectId=4638
% for recent versions of laprint.m.
%
% created by:           LaPrint version 3.16 (13.9.2004)
% created on:           27-Aug-2009 16:55:08
% eps bounding box:     15 cm x 7.8516 cm
% comment:              
%
\begin{psfrags}%
\psfragscanon%
%
% text strings:
\psfrag{s01}[t][t]{\color[rgb]{0,0,0}\setlength{\tabcolsep}{0pt}\begin{tabular}{c}Gain~[dB]\end{tabular}}%
\psfrag{s02}[b][b]{\color[rgb]{0,0,0}\setlength{\tabcolsep}{0pt}\begin{tabular}{c}$\Delta T$~[ps]\end{tabular}}%
%
% xticklabels:
\psfrag{x01}[t][t]{0}%
\psfrag{x02}[t][t]{0.1}%
\psfrag{x03}[t][t]{0.2}%
\psfrag{x04}[t][t]{0.3}%
\psfrag{x05}[t][t]{0.4}%
\psfrag{x06}[t][t]{0.5}%
\psfrag{x07}[t][t]{0.6}%
\psfrag{x08}[t][t]{0.7}%
\psfrag{x09}[t][t]{0.8}%
\psfrag{x10}[t][t]{0.9}%
\psfrag{x11}[t][t]{1}%
\psfrag{x12}[t][t]{0}%
\psfrag{x13}[t][t]{20}%
\psfrag{x14}[t][t]{40}%
\psfrag{x15}[t][t]{60}%
%
% yticklabels:
\psfrag{v01}[r][r]{0}%
\psfrag{v02}[r][r]{0.2}%
\psfrag{v03}[r][r]{0.4}%
\psfrag{v04}[r][r]{0.6}%
\psfrag{v05}[r][r]{0.8}%
\psfrag{v06}[r][r]{1}%
\psfrag{v07}[r][r]{0}%
\psfrag{v08}[r][r]{50}%
\psfrag{v09}[r][r]{100}%
\psfrag{a}[][t]{\textbf{a}}%
\psfrag{b}[][]{\textbf{b}}%
\psfrag{c}[][]{\textbf{c}}
\psfrag{d}[][]{\textbf{d}}%
% Figure:
\includegraphics[width=\textwidth]{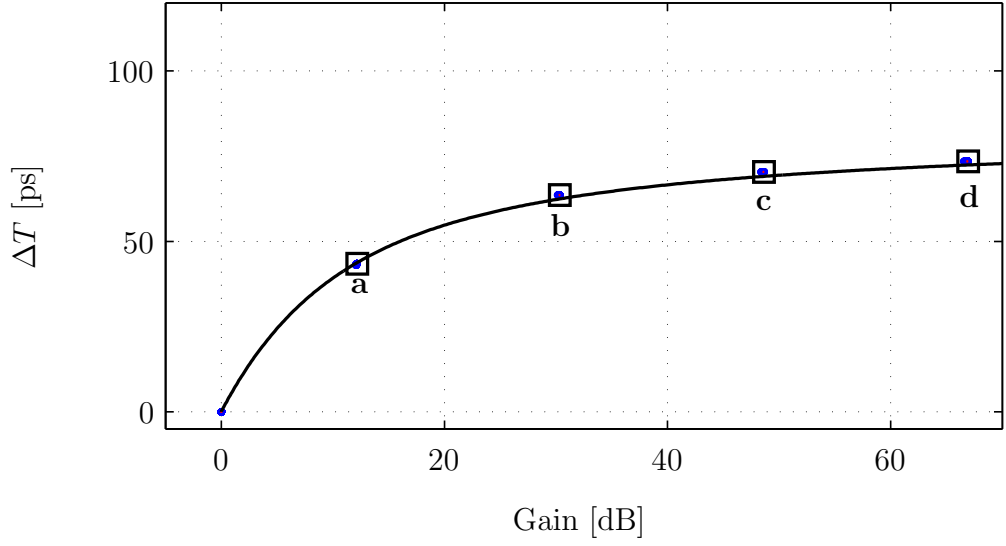}%
\end{psfrags}%
%
% End figure3.tex

  \end{center}
%  \centerline{\includegraphics[width=\textwidth]{figures/delayvsgainp1.jpg}}
  \caption{Same as fig. \ref{figus1}, but for a US fiber, $p=0.5$~m.}
\label{figus3}
\end{figure}

Last, the case of US is considered. 
Fig. \ref{figus2} shows a statistical set of realizations of a fiber with the same value of
$D$ as in fig. \ref{figus1}, but spun with a pitch $p=2$~m. We see that
the delay standard deviation is greatly reduced, and gets down to $5$~ps.
An impressive reduction of the standard deviation is obtained by further decreasing the pitch,
as shown in fig. \ref{figus3}, where $p=0.5$~m.  Note that there is no longer a
significant difference with respect to the case of an ideally isotropic fiber.

\begin{figure}
\begin{center}
  % This file is generated by the MATLAB m-file laprint.m. It can be included
% into LaTeX documents using the packages graphicx, color and psfrag.
% It is accompanied by a postscript file. A sample LaTeX file is:
%    \documentclass{article}\usepackage{graphicx,color,psfrag}
%    \begin{document}\input{muchi}\end{document}
% See http://www.mathworks.de/matlabcentral/fileexchange/loadFile.do?objectId=4638
% for recent versions of laprint.m.
%
% created by:           LaPrint version 3.16 (13.9.2004)
% created on:           14-May-2009 13:26:41
% eps bounding box:     15 cm x 7.6055 cm
% comment:              
%
\begin{psfrags}%
\psfragscanon%
%
% text strings:
\psfrag{s01}[t][t]{\color[rgb]{0,0,0}\setlength{\tabcolsep}{0pt}\begin{tabular}{c}$p$~[m]\end{tabular}}%
\psfrag{s02}[b][b]{\color[rgb]{0,0,0}\setlength{\tabcolsep}{0pt}\begin{tabular}{c}$(2 \mu)^{1/2}, \chi [\mathrm{m}^{-1}]$\end{tabular}}%
%
% xticklabels:
\psfrag{x01}[t][t]{0}%
\psfrag{x02}[t][t]{0.1}%
\psfrag{x03}[t][t]{0.2}%
\psfrag{x04}[t][t]{0.3}%
\psfrag{x05}[t][t]{0.4}%
\psfrag{x06}[t][t]{0.5}%
\psfrag{x07}[t][t]{0.6}%
\psfrag{x08}[t][t]{0.7}%
\psfrag{x09}[t][t]{0.8}%
\psfrag{x10}[t][t]{0.9}%
\psfrag{x11}[t][t]{1}%
\psfrag{x12}[t][t]{0}%
\psfrag{x13}[t][t]{0.5}%
\psfrag{x14}[t][t]{1}%
\psfrag{x15}[t][t]{1.5}%
\psfrag{x16}[t][t]{2}%
\psfrag{x17}[t][t]{2.5}%
\psfrag{x18}[t][t]{3}%
%
% yticklabels:
\psfrag{v01}[r][r]{0}%
\psfrag{v02}[r][r]{0.1}%
\psfrag{v03}[r][r]{0.2}%
\psfrag{v04}[r][r]{0.3}%
\psfrag{v05}[r][r]{0.4}%
\psfrag{v06}[r][r]{0.5}%
\psfrag{v07}[r][r]{0.6}%
\psfrag{v08}[r][r]{0.7}%
\psfrag{v09}[r][r]{0.8}%
\psfrag{v10}[r][r]{0.9}%
\psfrag{v11}[r][r]{1}%
\psfrag{v12}[r][r]{0}%
\psfrag{v13}[r][r]{ }%
\psfrag{v14}[r][r]{0.01}%
\psfrag{v15}[r][r]{ }%
\psfrag{v16}[r][r]{0.02}%
\psfrag{v17}[r][r]{ }%
\psfrag{v18}[r][r]{0.03}%
\psfrag{v19}[r][r]{ }%
\psfrag{v20}[r][r]{0.04}%
\psfrag{v21}[r][r]{ }%
\psfrag{v22}[r][r]{0.05}%
%
% Figure:
\centering{\includegraphics[width=\textwidth]{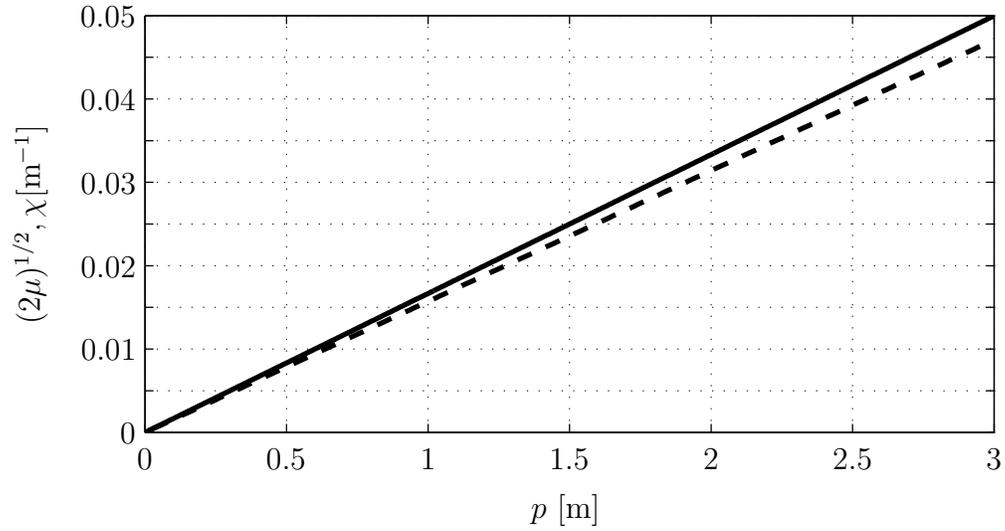}}
\end{psfrags}%
%
% End muchi.tex

  \end{center}
  \caption{Strength of the SM equivalent birefringence vector components, 
$\sqrt{2 \mu}$ (linear, dashed line) and $\chi$ (circular, solid line), as functions of the 
spin pitch $p$, as from eqs. \ref{muegamma} ($L_B=10$~m, $L_F=9$~m).}
\label{figmuegamma}
\end{figure}

This remarkable result can be explained by the fact that in US fibers, as the spin pitch decreases, 
the equivalent deterministic circular birefringence and the random linear birefringence 
decrease. This is shown in fig. \ref{figmuegamma}, where the strengths of
the random linear and deterministic circular components of the equivalent 
vector are calculated from eqs. \ref{muegamma}. It is then clear that US fibers behave similarly to
ideal isotropic fibers, when the spin pitch is short enough. The final outcome is that the pump and signal
SOPs remain almost parallel, for all possible SOPs launched at the fiber input.

\begin{figure}
\begin{center}
  % This file is generated by the MATLAB m-file laprint.m. It can be included
% into LaTeX documents using the packages graphicx, color and psfrag.
% It is accompanied by a postscript file. A sample LaTeX file is:
%    \documentclass{article}\usepackage{graphicx,color,psfrag}
%    \begin{document}\input{unspuncos}\end{document}
% See http://www.mathworks.de/matlabcentral/fileexchange/loadFile.do?objectId=4638
% for recent versions of laprint.m.
%
% created by:           LaPrint version 3.16 (13.9.2004)
% created on:           14-May-2009 16:29:42
% eps bounding box:     15 cm x 7.6179 cm
% comment:              
%
\begin{psfrags}%
\psfragscanon%
%
% text strings:
\psfrag{s01}[t][t]{\color[rgb]{0,0,0}\setlength{\tabcolsep}{0pt}\begin{tabular}{c}$f$~[GHz]\end{tabular}}%
\psfrag{s02}[b][b]{\color[rgb]{0,0,0}\setlength{\tabcolsep}{0pt}\begin{tabular}{c}$\langle \cos \theta_{p,s}\rangle$\end{tabular}}%
\psfrag{s06}[][]{\color[rgb]{0,0,0}\setlength{\tabcolsep}{0pt}\begin{tabular}{c} \end{tabular}}%
\psfrag{s07}[][]{\color[rgb]{0,0,0}\setlength{\tabcolsep}{0pt}\begin{tabular}{c} \end{tabular}}%
%
% xticklabels:

\psfrag{x12}[t][t]{-15}%
\psfrag{x13}[t][t]{-10}%
\psfrag{x14}[t][t]{-5}%
\psfrag{x15}[t][t]{0}%
\psfrag{x16}[t][t]{5}%
\psfrag{x17}[t][t]{10}%
\psfrag{x18}[t][t]{15}%
%
% yticklabels:

\psfrag{v12}[r][r]{0}%
\psfrag{v13}[r][r]{0.2}%
\psfrag{v14}[r][r]{0.4}%
\psfrag{v15}[r][r]{0.6}%
\psfrag{v16}[r][r]{0.8}%
\psfrag{v17}[r][r]{1}%
%
% Figure:
\centering{\includegraphics[width=\textwidth]{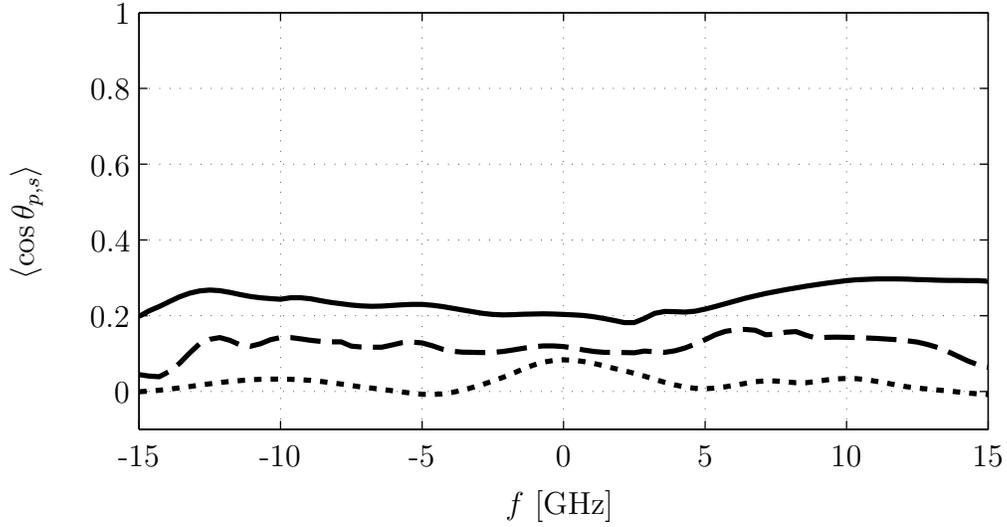}}%
\end{psfrags}%
%
% End unspuncos.tex

  \end{center}
  \caption{Alignment factor $cos(\theta_{p,s})$ between the output signal and pump SOPs in Stokes space,
as a function of the frequency detuning from signal carrier frequency for a LoBi fiber; dotted, 
dashed and solid curves are for $P_0=1,3,5$~W respectively.}
\label{figunspuncos}
\end{figure}

\begin{figure}
\begin{center}
  % This file is generated by the MATLAB m-file laprint.m. It can be included
% into LaTeX documents using the packages graphicx, color and psfrag.
% It is accompanied by a postscript file. A sample LaTeX file is:
%    \documentclass{article}\usepackage{graphicx,color,psfrag}
%    \begin{document}\input{spuncos}\end{document}
% See http://www.mathworks.de/matlabcentral/fileexchange/loadFile.do?objectId=4638
% for recent versions of laprint.m.
%
% created by:           LaPrint version 3.16 (13.9.2004)
% created on:           14-May-2009 16:44:05
% eps bounding box:     15 cm x 7.6179 cm
% comment:              
%
\begin{psfrags}%
\psfragscanon%
%
% text strings:
\psfrag{s01}[t][t]{\color[rgb]{0,0,0}\setlength{\tabcolsep}{0pt}\begin{tabular}{c}$f$~[GHz]\end{tabular}}%
\psfrag{s02}[b][b]{\color[rgb]{0,0,0}\setlength{\tabcolsep}{0pt}\begin{tabular}{c}$\langle \cos \theta_{p,s}\rangle$\end{tabular}}%
\psfrag{s06}[][]{\color[rgb]{0,0,0}\setlength{\tabcolsep}{0pt}\begin{tabular}{c} \end{tabular}}%
\psfrag{s07}[][]{\color[rgb]{0,0,0}\setlength{\tabcolsep}{0pt}\begin{tabular}{c} \end{tabular}}%
%
% xticklabels:
\psfrag{x12}[t][t]{-15}%
\psfrag{x13}[t][t]{-10}%
\psfrag{x14}[t][t]{-5}%
\psfrag{x15}[t][t]{0}%
\psfrag{x16}[t][t]{5}%
\psfrag{x17}[t][t]{10}%
\psfrag{x18}[t][t]{15}%
%
% yticklabels:
\psfrag{v12}[r][r]{0}%
\psfrag{v13}[r][r]{0.2}%
\psfrag{v14}[r][r]{0.4}%
\psfrag{v15}[r][r]{0.6}%
\psfrag{v16}[r][r]{0.8}%
\psfrag{v17}[r][r]{1}%
%
% Figure:
\centering{\includegraphics[width=\textwidth]{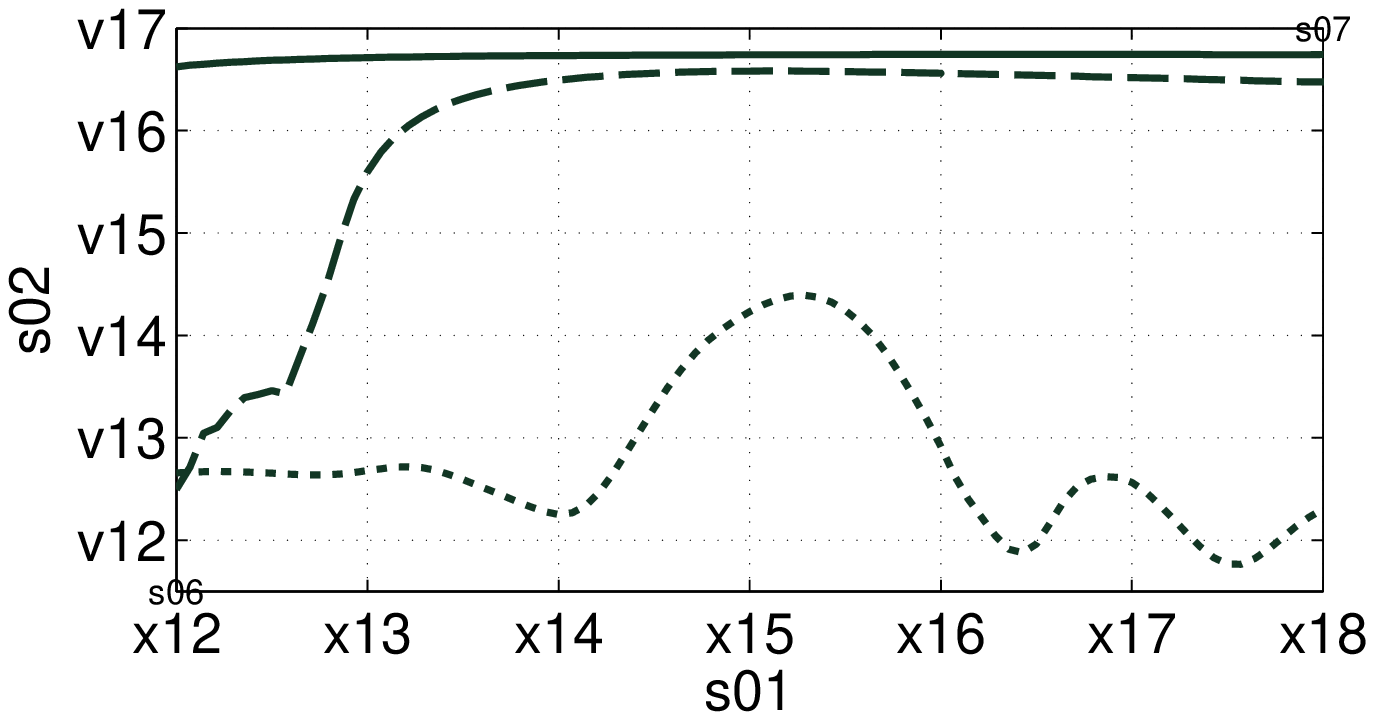}}%
\end{psfrags}%
%
% End spuncos.tex

  \end{center}
  \caption{Same as fig. \ref{figunspuncos}, but for a US fiber, $p=2$~m.}
\label{figspuncos}
\end{figure}

The enhanced parallelism is illustrated also by figs. \ref{figunspuncos} and \ref{figspuncos} , 
where the mean value of $cos(\theta_{p,s})$ ($\theta_{p,s}$ being the angle between the signal and 
pump SOPs in Stokes space, at the fiber output) is shown for the two cases (unspun and US), as a function 
of detuning from the signal carrier frequency.
Note that the alignment increases as the spin pitch decreses; moreover, it increses with
pump power. The latter effect is, once again, an indication of the nonlinear
polarization pulling effect that we mentioned in the previous section \cite{ZAD08OE,MAR08OE}.

%\begin{figure}
%\input{figures/figurePMbir.tex}
%  \caption{Effects of random coupling in a HiBi fiber (L=50 m, 
%providing $0.1293~\mathrm{ps}/\mathrm{m}$ 
%$D=0.073~\mathrm{ps}/\sqrt{\mathrm{km}}$ and $P_0=10$~W); each dot represents a statistical
%realizations of the fiber; the square corresponds to the isotropic case.}
%\label{figpmdist}
%\end{figure}

\section{Conclusions}
\label{conc}
The problem of how to control the SOP of the signal
and pump, for slow and fast light in narrow-band, Raman-assisted, optical parametric amplification, 
has been studied theoretically and numerically. It was shown that standard,
unspun telecommunication fibers
can exhibit delay fluctuations, caused by polarization mode dispersion,
that are too large in order to yield reliable slow and fast light
effects for practical applications.

To mimic an ideal isotropic fiber, two options have been explored: high birefringence
fibers, and unidirectionally spun fibers. 

For HiBi polarization maintaining fibers, the typical achievable polarization crosstalk
ratio is sufficiently good to guarantee that an isotropic-fiber like delay is obtained.
The effects of practical signal input misalignments are also under control,
thanks to the nonlinear polarization pulling
effect, which attracts the signal SOP towards the direction yielding maximum gain.
Furthermore, small input pump misalignents, to be expected in practice,
are not affecting the delay to a significant level.

Finally, unidirectionally spun fibers were considered. For fast spinning
($0.5$~m), they have been shown to behave essentially like ideal isotropic fibers.
In fact, compared to unspun fibers, the polarization mean alignment is
highly enhanced along the entire length of the fiber.
An additional advantage of spun fibers is that they maintain aligned {\it any} state of polarization, so the only care they require is to align signal
and pump at the fiber input along the same direction.

From a practical viewpoint, high birefringence commercial fibers are readily
available, though customization might be necessary for tayloring the dispersion
properties to match the typical wavelengths of pump sources.
As for unidirectionally spun fibers, spin pitches down to a few mm are feasible 
with present technologies. Therefore, such fibers could indeed be
the most suitable tools for implementing reliable, wideband, slow and fast
light devices.

\section{Acknowledgments}
The research leading to these results has received funding from the European
Community's Seventh Framework Programme (FP7/2007-2011). under grant agreement n 219299 "GOSPEL". 
Part of this work was done within the agreement between the University of Padova and
ISCOM, Rome, Italy. Part was supported by the Italian Ministry of Foreign
Affairs (Direzione
Generale per la Promozione e la Cooperazione Culturale).


%\appendix
\section{Appendix}
\label{appendice}
The nonlinear operators of eqs.~\ref{fopaeq} are defined accordingly to the following expressions:
\begin{eqnarray}
{\cal{S}}_p &=&  j \dfrac{\gamma^{NR}_p}{3} \left[2\langle A_p|A_p\rangle \mathbf{I}+|A_p^\ast\rangle \langle A_p^\ast|\right]+\nonumber\\%
&&\,\,+j \left[\left(\chi_{p,1212}^{R}(0)+\chi_{p,1122}^{R}(0)\right)\langle A_p|A_p\rangle \mathbf{I}+\chi_{p,1221}^{R}(0)|A_p^\ast\rangle \langle A_p^\ast|\right]\\
{\cal{X}}_{s,i} &=&  2j \dfrac{\gamma^{NR}_{s,i}}{3} \left[\langle A_p|A_p\rangle \mathbf{I}+|A_p\rangle \langle A_p|+|A_p^\ast\rangle \langle A_p^\ast|\right]\\
{\cal{F}}_{s,i} &=&  j \dfrac{\hat{\gamma}^{NR}_{s,i}}{3} \left[\langle A_p^\ast|A_p\rangle \mathbf{I}+2|A_p\rangle \langle A_p^\ast|\right]+\nonumber\\%
&&\,\,+j \left[\left(\hat{\chi}_{s,i,1122}^{R}(\Omega)+\hat{\chi}_{s,i,1221}^{R}(\Omega)-\hat{\chi}_{s,i,1212}^{R}(\Omega)\right)\langle A_p^\ast|A_p\rangle \mathbf{I}+\right.\nonumber\\%
&&\,\,+\left(2\hat{\chi}_{s,i,1212}^{R}(\Omega)+\hat{\chi}_{s,i,1122}^{R}(\Omega)+\hat{\chi}_{s,i,1221}^{R}(\Omega)\right.\nonumber\\%
&&\,\,-\left.\left.\hat{\chi}_{s,i,1122}^{R}(0)-\hat{\chi}_{s,i,1221}^{R}(0)\right)|A_p\rangle \langle A_p^\ast|\right]\\%
{\cal{R}}_{s,i} &=&  j \left[2\chi_{s,i,1122}^{R}(\Omega)\langle A_p|A_p\rangle \mathbf{I}+2\chi_{s,i,1212}^{R}(\Omega)|A_p\rangle \langle A_p|+\right.\nonumber\\%
&&\,\,\left.+2\chi_{s,i,1221}^{R}(\Omega)|A_p^\ast\rangle \langle A_p^\ast|\right]%
\end{eqnarray}
where%
\begin{eqnarray}
\gamma_p^{NR} &= &\frac{2\pi \omega_p^2}{c^2 \beta_p A_p}\chi_{1122}^{NR},\\%
\gamma_{s,i}^{NR} &= &\frac{2\pi \omega_{s,i}^2}{c^2 \beta_{s,i} A_{s1,i1}}\chi_{1122}^{NR},\\%
\hat{\gamma}_{s,i}^{NR} &= &\frac{2\pi \omega_{s,i}^2}{c^2 \beta_{s,i} A_{s2,i2}}\chi_{1122}^{NR},%
\end{eqnarray}
with%
\begin{equation}
\chi_{1111}^{NR}=3\chi_{1212}^{NR}=3\chi_{1122}^{NR}=3\chi_{1221}^{NR},%
\end{equation}
and%
\begin{eqnarray}
\chi_{p,klmn}^{R}(\Omega) &=& \frac{2\pi \omega_p^2}{c^2 \beta_p A_p}\chi_{klmn}^{R}(\Omega),\\%
\chi_{s,i,klmn}^{R}(\Omega)& =& \frac{2\pi \omega_{s,i}^2}{c^2 \beta_{s,i} A_{s1,i1}}\chi_{klmn}^{R}(\Omega),\\%
\hat{\chi}_{s,i,klmn}^{R}(\Omega)& =& \frac{2\pi \omega_{s,i}^2}{c^2 \beta_{s,i} A_{s2,i2}}\chi_{klmn}^{R}(\Omega),%
\end{eqnarray}
with%
\begin{equation}
\chi_{1111}^{R}(\Omega)= \chi_{1212}^{R}(\Omega)+\chi_{1122}^{R}(\Omega)+\chi_{1221}^{R}(\Omega).
\end{equation}
and %
\begin{equation}
\Omega= \omega_p-\omega_i=\omega_s-\omega_p.
\end{equation}

The expression of the nonresonant components of the nonlinearity coefficient $\chi_{klmn}^{NR}$ and of the resonant
components of the nonlinear Raman susceptibility $\chi_{klmn}^{R}(\Omega)$ can be found in \cite{TRIL94JOSAB}.
The effective areas $A_p$ and $A_{s1,i1}$, $A_{s2,i2}$ read
\begin{eqnarray}
A_p &=& \dfrac{\langle f_p^2\rangle^2}{\langle f_p^4\rangle},\\%
A_{s1,i1}&=&\dfrac{\langle f_p^2\rangle\langle f_{s,i}^2\rangle}{\langle f_p^2 f_{s,i}^2\rangle},\\%
A_{s2,i2}&=&\dfrac{\langle f_p^2\rangle\sqrt{\langle f_{s}^2\rangle}\sqrt{\langle f_{i}^2\rangle}}{\langle f_p^2 f_s f_i\rangle},%
\end{eqnarray}
where here angle brackets stand for integrals over the trasversal modal profiles $f_{j}=f_{j}(x,y)$ ($j=p,s,i$).
In the numerical simulation of this paper it has been assumed $A_p=A_{s1,i1}=A_{s2,i2}$.

\end{document}